\documentclass[12pt]{iopart}

\usepackage{graphicx}
\usepackage{dcolumn}
\usepackage{bm}
\usepackage{amssymb}

\usepackage{color}
\usepackage{ulem}

\begin{document}

\title {The magnetic monopole and the separation between fast and slow magnetic degrees of freedom}
\author{J.-E.
Wegrowe}
\address{LSI, Ecole Polytechnique, CEA, CNRS,  Universit\'e Paris-Saclay, 91128 Palaiseau Cedex, France}
\author{E. Olive}
\address{GREMAN, UMR 7347, Universit\'e Fran\c cois Rabelais-CNRS, Parc de Grandmont, 37200 Tours, France}
 \ead{jean-eric.wegrowe@polytechnique.edu}
  
\date{\today}

\begin{abstract}
The Landau-Lifshitz-Gilbert (LLG) equation that describes the dynamics of a macroscopic magnetic moment finds its limit of validity at very short times. The reason for this limit is well understood in terms of separation of the characteristic time scales between slow degrees of freedom (the magnetization) and fast degrees of freedom. The fast degrees of freedom are introduced as the variation of the angular momentum responsible for the inertia. In order to study the effect of the fast degrees of freedom on the precession, we calculate the geometric phase of the magnetization (i.e. the Hannay angle) and the corresponding magnetic monopole. In the case of the pure precession (the slow manifold), a simple expression of the magnetic monopole is given as a function of the slowness parameter, i.e. as a function of the ratio of the slow over the fast characteristic times. 
\end{abstract} 

\maketitle

Recently, important efforts have been devoted to both the reformulation of well known effects and to the description of new phenomena by means of the geometric phase (the quantum Berry phase \cite{Berry,Aharonov} or the classical Hannay angle \cite{Hannay,Berry2}), in particular in relation to spin systems \cite{Bruno}. 

The geometric phase is indeed an efficient tool that allows the essential physics to be extracted from a complex system, in which gauge invariance plays a fundamental role (e.g. in terms of ``curl forces'' \cite{Berry3} or ``equilibrium currents'' \cite{Sonin}). 
An important application can be found for electronic transport in ferromagnets, typically for the anomalous Hall effect \cite{AHE}, or for the recent developments about electronic devices that exploit spin-orbit interactions \cite{Xiao,Nagaosa, Nagasawa,Bruno04,Shibata, Barnes05,Kurebayashi}. The geometric phase appears to be also a necessary tool for the description of the transport of magnetic moments or spins \cite{AharonovCasher,Aharonov2,Hertel04,Hertel}, or for the description of magnetic excitations traveling throughout chiral structures \cite{Tserkovnyak,Onose,Murakami}.

In the above mentioned cases, the magnetic configuration is not always at equilibrium. Instead, a transport effect occurs also inside the magnetic or spin configuration space, at each point of the real space. The corresponding magnetization dynamics are described by the well-known Landau-Lifshitz-Gilbert equation (LLG) \cite{Landau,Gilbert,Gilbert2,Baryarkhtar}. If one consider both the transport throughout the usual configuration space and inside the magnetization space, the set of possible magnetic excitations is extraordinarily rich and complex \cite{Bertotti,Lakshmanan,Pylyovskyi,Robbins1}. Even if one consider only the case of uniform magnetization (no space variable), the LLG equation already describes a wide variety of effects, including ferromagnetic resonance and rotational brownian motion in a field of force \cite{Siegmann,Miltat,Brown,Coffey,Cugliandolo,Skrotskii,Maamache}.

Furthermore, recent investigations suggest that, at the ultra-fast regime, the LLG equation should be generalized with considering inertial terms \cite{inertia,inertia1,inertia2,inertia3,inertia4,inertia5,Thonig,Li}. The goal of the present work is to investigate the inertial regime for the uniform magnetization with the help of the geometric phase.  In this context, we focus our attention to the connection between three fundamental concepts; the {\it geometric phase} of the magnetization, the {\it magnetic monopole}, and the {\it inertial regime} of the magnetization. The three concepts are coupled because the dynamics of a magnetic dipole are composed of both fast and slow dynamics, and the geometric phase is an efficient tool for the study of the separation of time-scales between slow and fast degrees of freedom \cite{Aharonov3,Robbins3}. The influence of the fast variables on the slow motion is treated in perturbation expansions \cite{VanKampen} in which the ratio of small and fast time scales define a slowness parameter, and the successive terms are interpreted as reaction forces of the fast variables on the slow motion \cite{BerryShukla,SpinTop}\\

The magnetization $\vec M$ of a uniformly magnetized body is usually defined as a magnetic dipole. The description of the dynamics of a classical magnetic dipole is however still problematic today \cite{Griffiths}. Ampere's magnetic dipole is defined by an electric charge that is moving {\it at high speed} about a microscopic ``loop'', typically an atomic orbital. This simple model allows the gyromagnetic relation to be derived : the magnetization $\vec M$ of the magnetic dipole then follows the angular momentum  $\vec L$ of the electric carrier, with the relation $\vec M = \gamma \vec L$ where $\gamma = gq/(2m)$ is the gyromagnetic ratio ($m$ is the mass and $q$ is the electric charge of the electric carrier, and $g$ is the Land\'e factor).

If a static magnetic field $H_z$ (oriented along $\vec e_z$) is applied, the magnetization precesses at the Larmor angular velocity $\Omega_L$ around the axis defined by $\vec e_z$. In other terms, a {\it slow motion} (precession) is added to the {\it fast motion} (moving electric carrier) that defines the magnetic dipole. In the absence of dissipation
 the dynamics of the dipole are reduced to a simple precessional motion. However, this reduction is valid only if the velocity of the electric charge is much higher than the precession velocity, i.e. if the typical time-scales are well separated.

Indeed, if the Larmor angular velocity is high enough and becomes of the same order as the angular velocity of the electrical carrier moving in the loop, the Amperian magnetic dipole $\vec M$ is no longer defined by a simple expression (the exact trajectory of the punctual electric carrier should be taken into account instead of averaging over the loop) \cite{LLCourse,McDonald}. 

However there is an other way to define a magnetic dipole, namely the {\it Gilbert's dipole} (according to D. J. Griffiths, the Gilbert dipole is a double monopole \cite{Griffiths2}). In our non-relativistic context, the Gilbert magnetic dipole is defined by its dynamical properties, based on the mechanical analogy with the spinning top \cite{Gilbert}. This mechanical approach allowed T. H. Gilbert to derive the well-known Landau-Lifshitz-Gilbert equation (LLG), providing that the first two principal moments of inertia vanish $I_1 = I_2 = 0$, but not the third one $I_3 \ne 0$ \cite{Gilbert2}. This ad-hoc assumption is related to the electrodynamic limitation of the Amperian magnetic dipole mentioned above.

In this context, fast degrees of freedom have been taken into account as inertial variables (so that $I_1= I_2 \ne 0$) by enlarging the configuration space to the corresponding phase space, i.e. including the angular momentum. The corresponding generalized LLG equation then contains a supplementary term proportional to the second time-derivative of the magnetization \cite{inertia,inertia1,inertia2,inertia3,inertia4,inertia5,Thonig}.\\

In the present work, we show that the Hannay angle and the corresponding magnetic monopole are able to describe, in the adiabatic limit, the transition from the usual precession to more complex dynamics containing the inertial effects. The analysis follows the method recently proposed by M. V. Berry and P. Shukla in Ref.\cite{SpinTop} for the study of the spinning top. Within this approach, the dynamics of the magnetization are interpreted as the reaction of the fast dynamics on the slow. A simple analytical result is obtained by reducing the phase space to the slow manifold. 

The paper is composed as follows. Section 1 below is devoted to the mechanical definition of the adiabatic Gilbert dipole without taking into account the fast degrees of freedom. Section 2 describes the adiabatic kinetic equation. The geometric phase is presented in section {\it 2.3}, and the corresponding magnetic monopole is described in section {\it 2.4}. Section 3 studies the effect of the fast degrees of freedom. In particular, the calculation of the adiabatic dynamics of the magnetization that includes inertia is presented in section {\it 3.1}, and the calculation of the geometric phase with inertia is given in section {\it 3.2}. The case of the pure precession is studied in section {\it 3.3}, and the corresponding magnetic monopole is given in section {\it 3.4}. The conclusion is proposed in Section {\it 4}.

\section{The Gilbert magnetic dipole} 

Gilbert's mechanical model is sketched in Fig.~1. A rigid cylindrical
stick of length $M_s$, with one end fixed at the origin, is pointing in a direction described by the angles $\theta$ and $\varphi$. The magnetization is aligned along the effective magnetic field $H_z$ at equilibrium.  Due to the application of a vertical force oriented along the $z$ axis, the stick is precessing around the vertical axis at angular velocity $\dot \varphi$. The magnetic energy is $V^F = - \vec M. \vec H$ where $\vec H$ is the effective field and $\vec M = M_s \vec e_3$ is the magnetization ($\vec e_3$ is the unit vector defined in Fig.1). 
Furthermore, the stick is spinning around its own symmetry axis at angular velocity $\dot \psi$. This motion corresponds to the rotation of the electric carrier of Ampere's dipole (see below). The phase space of this rigid rotator is
defined by the angles $\{ \theta, \varphi, \psi \}$ and the three components of the associated angular momentum $\vec L$.
The relation between the angular momentum 
and the angular velocity $\vec \Omega$ is $\vec L =  \bar{\bar{I}} 
\vec \Omega$, where $\bar{\bar{I}}$ is the inertia tensor. 

 In the rotating frame, or body-fixed frame
$\{ \vec e_1, \vec e_2, \vec e_3 \}$, the inertial tensor 
is reduced to the principal moments of inertia $\{ I_1,I_2,I_3 \}$. 
The symmetry of revolution of the spinning stick imposes furthermore that $I_1=I_2$ :
\begin{equation}
\bar{\bar{I}} = \left( \begin{array}{ccc}
 I_1 &  0 & 0 \\
                       0 & I_1 & 0 \\
                       			0 & 0 & I_3
\end{array} \right).
\label{MatrixL0}
\end{equation}
In the fixed body frame, the angular velocity reads (see Fig. 1) :
\begin{eqnarray}
\Omega_1 &=  \dot \varphi  \, \sin \theta \, \sin \psi + \dot \theta \, \cos \psi, \\
\Omega_2  &= \dot \varphi  \, \sin \theta \, \cos \psi - \dot \theta \, \sin \psi, \\
\Omega_3 & = \dot \varphi  \, \cos \theta \  + \dot  \psi.
\label{Omega0}
\end{eqnarray}

 \begin{figure} [h!]
   \begin{center}
   \begin{tabular}{c}
 \includegraphics[height=6cm]{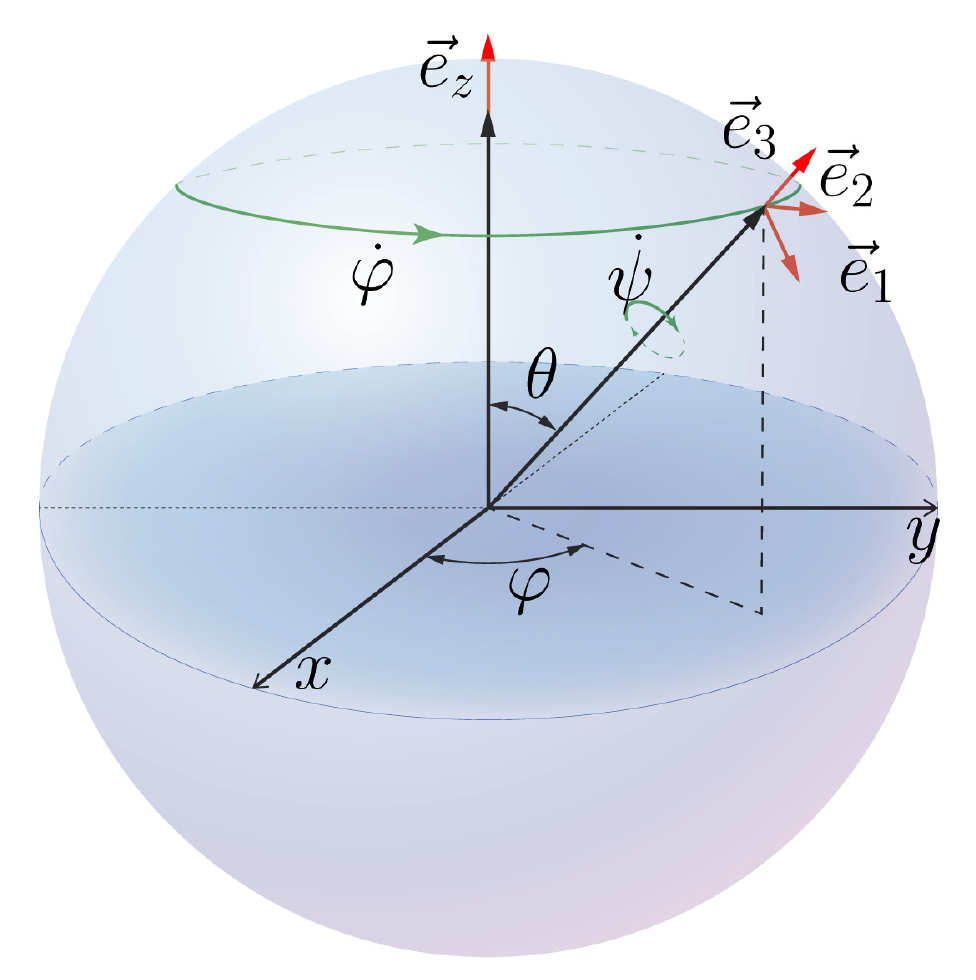}
   \end{tabular}
   \end{center}
   \caption[SpinTop]
{ \label{fig:SpinTop} Illustration of the magnetomechanical analogy of a 
spinning stick that precesses around the $z$ axis. The coordinates of the stick in the space-fixed frame are parameterized by the angles $ ( \theta,  \varphi, \psi )$ and the radius of the sphere is $M_s$. The body-fixed frame, denoted $\{\vec e_1,\vec e_2, \vec e_3\}$, is spinning with angular velocity $\dot \psi$ and is precessing around $\vec e_z$ with angular velocity $\dot \varphi$.}
   \end{figure}

The kinetic equation is obtained from the angular velocity : for any vector $\vec M$ of constant modulus carried with the rotating body, we have :

\begin{equation}
\frac{d\vec M}{dt} = \vec \Omega \times \vec M
\label{kinetic0}
\end{equation}

\section{The kinetic equation}

 \subsection{Gyromagnetic relation}
Let us start with Gilbert's hypothesis of vanishing inertia \cite{inertia} : $I_1 = I_2 \rightarrow 0$ so that $L_1 = L_2 \rightarrow 0$. However, we have $L_3 = I_3 \Omega_3 \ne 0$. 
Since $\vec L = L_3 \vec e_3$, the conservation of angular momentum $\vec L$ imposes $L_3$ constant (this is also valid in the case of damping \cite{inertia}). Without loss of generality, we can define the modulus of the vector $\vec M$ with the help of the constant $\gamma$, such that 
\begin{equation}
M_s =  \gamma L_3 =  \gamma \Omega_3 I_3
\label{Gyromag}
\end{equation}
where $\gamma$ defines the well-known gyromagnetic ratio. 
 
 \subsection{Precession equation without damping}
 The effective magnetic field is defined by the canonical relation $\vec H = - \vec \nabla_{\Sigma} V^F$ where $\vec \nabla_{\Sigma} \equiv \partial / \partial \vec M$ is the gradient defined on the configuration space $\Sigma$ (which is the surface of the sphere of radius $M_s$). The torque exerted on the system is defined by the vectorial product $\vec \Gamma = \vec M  \times  (- \vec \nabla_{\Sigma} V^F)=  \vec M \times \vec H$. By convention,  we defined the direction $\vec e_z$ along the effective field $\vec H = H_z \vec e_z$. The third Newton's law $d \vec L/dt = \vec \Gamma$ gives then the kinetic equation of the magnetization :
\begin{equation}
\frac{d \vec{e_3}}{dt} = -\frac{M_sH_{z}}{L_3} (\vec e_3 \times \vec e_z).
\label{LLG}
\end{equation}
According to the gyromagnetic relation Eq.(\ref{Gyromag}), we have $M_sH_{z}/L_3 = \gamma H_{z}$ and equation Eq.(\ref{LLG}) is nothing but the well-known equation of the precession of the magnetization without damping : $\frac{d \vec{M}}{dt} = \gamma (\vec M \times \vec H)$.
Furthermore, since $\frac{d \vec{e_3}}{dt} = \vec \Omega \times \vec e_3$, the kinetic equation reads $\vec \Omega \times \vec e_3 = \frac{M_sH_{z}}{L_3} (\vec e_3 \times \vec e_z)$. Inserting the precession angular velocity $\Omega_{\varphi} = \dot \varphi$, we have :
\begin{equation}
\Omega_{\varphi} = \dot \varphi=  \frac{M_sH_{z}}{L_3} = -\gamma H_{z},
\label{Larmor}
\end{equation}

which is the definition of the Larmor angular velocity, as expected for a precessing magnetic moment.

\subsection{The geometric phase}
\label{Hannay}

 The geometric phase is the phase difference acquired over the course of a precession loop.
The precession time $t_s$ (i.e. the slow characteristic time of our problem) is the time at which the axis $\vec e_3$ is rotating one cycle around the axis $\vec e_z$, i.e. such that  $2 \pi = \int_0^{t_s} \left \vert \Omega_{\varphi}\right  \vert dt$. 

According to Eq.(\ref{Larmor}), the precession time is given by : 
\begin{equation}
t_s =  \left \vert \frac{2 \pi }{\Omega_{\varphi}} \right  \vert =    \frac{2 \pi}{\gamma H_z}.
\label{Precession_Time}
\end{equation}

 We can now give the expression of the number $ \Delta \Psi_0$ (the subscript $0$ stands for the non-inertial approximation) of rotation around the $\vec e_3$ axis (spinning rotation) during the time of a precession of the same axis around $\vec e_z$. According to the relation $\dot \psi = \Omega_{3}  - \dot \varphi cos \theta$, we have : 
\begin{equation}
\Delta \Psi_0 = \int_0^{t_s} \left ( \Omega_{3}  - \dot \varphi cos \theta \right ) dt =  \left ( \Omega_{3}  - \dot \varphi cos \theta \right )t_s = 2 \pi \left (\frac{M_s}{\gamma^2 I_3 H_z} + cos \theta \right),
\label{Deltapsi1}
\end{equation}
where we used Eq.(\ref{Precession_Time}), Eq.(\ref{Gyromag}), and the expression of the Larmor angular velocity $\Omega_{\varphi} = \dot \varphi = -\gamma H_z$. Anticipating over the next Section, we introduce the ``slowness parameter'' $G$ defined as the dimensionless angular momentum $L_3=I_3\Omega_3$ scaled with the angular momentum $\sqrt{I_1M_sH}$, i.e. the ratio of the slow over the fast angular momentum, or equivalently of the slow over the fast time-scale : 
\begin{equation}
G =  \frac{L_3}{\sqrt{I_1 M_s H_z}} =  \frac{1}{\gamma H_z} \sqrt{\frac{M_s H_z}{I_1}} \equiv \frac{1}{2 \pi} \frac{t_s}{t_0}.
\label{SlowG}
\end{equation}
 The last term in the right-hand side of Eq.(\ref{SlowG}) defines the fast characteristic time $t_0 = \sqrt{I_1/(M_s H_z)}$ of the motion. 

The expression of $\Delta \Psi_0$ now reads :
\begin{equation}
\Delta \Psi_0 =  2 \pi \left ( \frac{I_1}{I_3} G^2 + cos \theta \right)
\label{DeltapsiG}
\end{equation}

The first term in the right hand side can be defined as the dynamical angle, while the second term $2\pi cos \theta $ can be defined as the geometric phase (see however the discussion in reference \cite{SpinTop}). 
Note that the factor $(I_1/I_3) G^2 = \Omega_3/ \gamma H_z $ also defines a time ratio $t_s/t_f$, where $t_f=2 \pi /\Omega_3$ is another possible fast characteristic time of the movement. This parameter will be discussed below.
The expression Eq.(\ref{DeltapsiG}) is completed in Section IV below, in the case of inertia, with an expansion as a series of power of $cos \theta$.

\subsection{The magnetic monopole}

From the viewpoint of the geometric phase, the Gilbert's magnetic dipole is defined by the two magnetic monopoles $\pm B_{eff}$ that radiate from the center of a sphere of radius $R$ through both north (+) and south (-) hemispheres. The parameter $R$ is defined by Ampere's magnetic dipole $\vec M = \gamma \vec L $ that is generated by the electric carrier of charge $q$ and mass $m$ rotating inside the loop of radius $R$. The phase $\Delta \Psi_0$ then allows to link the mechanical definition of Gilbert's magnetic dipole to Ampere's magnetic dipole. 
 If we define the {\it radial field} $\vec B_{eff} = B_{eff} \, \vec e_3$ by a potential vector $\vec A = \vec{rot} (\vec B_{eff})$, the circulation of $\vec A$ around a closed loop of radius $R$ defines a phase \cite{Berry}
\begin{equation}
\Delta \Psi_0 \equiv \oint \vec A . d\vec l \equiv \int \int\vec B_{eff} . \vec {dS} =  \pi R^2 B_{eff},
\label{Deltapsi2}
\end{equation}
which is the geometric phase calculated above. Eq.(\ref{Deltapsi2}) and Eq.(\ref{Deltapsi1}) gives the expression of $B_{eff}$ :
\begin{equation}
B_{eff}= \frac{2}{R^2} \left ( \frac{M_s}{\gamma^2 I_3 H_z} + cos \theta \right)
\label{MonopoleMag}
\end{equation}

On the other hand, in the framework of the Ampere's model of the ``molecular currents'', a microscopic magnetic moment is defined by the {\bf Bohr magneton} $M_s = \mu_B = \gamma \hbar$ generated by an electron of mass $m$ and charge $q$ moving in a loop of Bohr radius $R$. The gyromagnetic ratio is $\gamma = q/(2m)$ and the moment of inertia associated to the loop of radius $R$ is $I_3 = m R^2$. 
Furthermore, the flux $\Phi_0 = \int H_z . dS $ of the external magnetic field (by convention along $Oz$) $\vec H = H_z \vec e_z$  through the microscopic hemisphere of radius $R$ is also quantified, with the well-known quantized flux : 
\begin{equation}
\Phi_0 = H_z  \pi R^2 cos \theta = \frac{h}{q}
 \label{FluxQ}
 \end{equation}
Equation (\ref{MonopoleMag}) then reads :
\begin{equation}
B_{eff}= \frac{4 cos \theta}{R^2} 
\label{MonopoleSpin}
\end{equation}
This expression defines the classical counterpart of the magnetic monopole \cite{Berry,Hannay,Berry2,Holstein,Garg}. Note that the corresponding geometric phase Eq.(\ref{DeltapsiG}) reduces to : $\Delta \Psi_0 =  4 \pi cos \theta$.

\section{The effect of inertia}
 
 \subsection{Inertial equation of the magnetization without damping}

The scalar gyromagnetic relation Eq.(\ref{Gyromag}) used above in the framework of the mechanical (or Gilbert's) model of the magnetic dipole coincides with the usual vectorial definition $\vec M = \gamma \vec L$ of the gyromagnetic relation if the inertial effects are neglected $I_1 = I_2 = 0$. If we take into account inertial effects, $I_1 = I_2 \ne 0$, the gyromagnetic relation $\vec M = \gamma \vec L$ is no longer valid in this form. The generalized equation is obtained, by cross- multiplication of Eq.(\ref{kinetic0}) with the vector $\vec M = M_s \vec e_3$. 
 
 \begin{equation}
 \vec \Omega = \frac{\vec M}{M_s^2} \times \frac{d \vec M}{dt}  + \Omega_3 \vec e_3,
 \label{Omega}
 \end{equation}
 or :
 
  \begin{equation}
 \vec L =  \frac{I_1}{M_s^2} \left ( \vec M \times \frac{d \vec M}{dt} \right ) + L_3 \vec e_3.
 \label{AngularMoment}
 \end{equation}
 Newton's law $d \vec L/dt = \vec M \times \vec H_{eff}$ becomes, with the constant $L_3 = M_s/\gamma$ :
 
   \begin{equation}
\frac{d \vec e_3}{dt} =  \gamma H_{z} \,  \vec e_3 \times \left ( \vec e_z - t_0^2 \, \frac{d^2 \vec e_3}{dt^2} \right).
 \label{ILLG}
 \end{equation}
 
 where the characteristic time $t_0 = \sqrt{I_1/M_s H_{z} }$ has already been introduced in Eq.(\ref{SlowG}). Equation (\ref{ILLG}) generalizes Eq.(\ref{LLG}) with the inertial term ($I_1 \ne 0$). This equation is the adiabatic limit (i.e. without damping) of the inertial LLG presented in previous studies \cite{inertia}.
  
It is convenient to rewrite Eq.(\ref{ILLG}), with the dimensionless time $\tau = t/(t_0)$ and the slowness parameter $G$ (both defined in Eq.(\ref{SlowG})). The equation of motion Eq.(\ref{ILLG}) takes the following vectorial form :

   \begin{equation}
\frac{d \vec e_3}{d \tau} = \frac{1}{G} \left ( \vec e_3 \times \vec e_z - \vec e_3 \times \frac{d^2 \vec e_3}{d \tau^2} \right)
 \label{ILLG2}
 \end{equation}

Eqs. (\ref{ILLG2})  becomes
\begin{eqnarray}
\label{ILLG3}
\theta'' & = & -G\varphi' \sin\theta + \varphi'^2 \sin\theta \cos\theta  - \sin\theta 
 \\
\varphi'' \sin\theta & = & G\theta'  - 2\varphi' \theta' \cos\theta
\nonumber
\end{eqnarray}
where 
$\theta'=d\theta/d \tau,\ \theta''=d^2\theta/d \tau^2,\ \varphi'=d\varphi/d \tau,\ \phi''=d^2\phi/d \tau^2$.

This equation is the dynamical equation of the magnetization generalized to inertial effects (in the absence of damping). These equations allow the adiabatic movement to be studied below in terms of the geometric phase. The generalized equation including Gilbert damping has been studied in previous reports \cite{inertia1,inertia3,inertia5}. 

\subsection{The geometric phase with inertia}
The number $ \Delta \Psi$ of rotation around the $\vec e_3$ axis performed by the magnetization vector during the (dimensionless) time $\tau_s=t_s/(2 \pi t_0)$ of one precession is : 
\begin{equation}
\Delta \Psi = \int_0^{\tau_s}  \Psi' d \tau=  \int_0^{\tau_s} \left(\widetilde \Omega_3 -\varphi' \cos\theta\right) d \tau,
\label{psi1}
\end{equation}

where $\widetilde\Omega_3$  is the dimensionless angular velocity $t_0 \Omega_3$. Due to the conservation of the angular momentum component $L_3$, $\Omega_3$ is constant which implies
\begin{eqnarray}
\nonumber
\Delta \Psi & = & \widetilde\Omega_3 \tau_s - \int_0^{\tau_s} \varphi'(\tau) \cos\theta(\tau) d \tau \\
 & = & \widetilde\Omega_3 \tau_s -2\pi+ \int_0^{\tau_s} \varphi'(\tau) \left(1-\cos\theta(\tau) \right) d \tau
\label{psi2}
\end{eqnarray}
The Hannay angle $\Delta \psi_H$ is
\begin{equation}
\Delta \psi_H= \int_0^{\tau_s} \varphi'(\tau) \left(1-\cos\theta(\tau) \right) d \tau
\end{equation}
which is the solid angle swept by the axis in one precession cycle.

\subsection{Pure precession : an exact solution}
Following Ref.\cite{SpinTop} we seek for the slow manifold, {\it i.e.}  the set of initial conditions in the phase space for which the particular solution of the equations of motion Eqs.(\ref{ILLG3}) corresponds to pure precession, which means precession in the absence of nutation. It therefore corresponds to $\theta'=0$, from which inserted in  Eq.(\ref{ILLG3}a) gives 
\begin{eqnarray}
\label{GPurePrecession}
G \varphi' =\varphi'^2 \cos\theta-1
\end{eqnarray}
The dynamics of pure precession therefore give two corresponding precessional velocities, a slow one $\varphi'_-$ and a fast one $\varphi'_+$, which are given by
\begin{eqnarray}
\label{PurePrecessions}
\varphi'_{\pm}=\frac{G}{2\cos\theta}\left(1\pm \sqrt{1+\frac{4\cos\theta}{G^2}}\right)
\end{eqnarray}

The square root in this equation shows that the pure precession requires $\cos\theta > -G^2/4$. Therefore, pure precession without nutation is possible for $\left\vert G \right\vert>2$ for any inclination angle $\theta$, whereas for $\left\vert G \right\vert<2$, pure precession is only possible for inclination angles such that $\cos\theta > -G^2/4$.
We now consider the slow precession velocity $\varphi'_-$ given by Eq.(\ref{PurePrecessions}).
For such slow pure precession it is possible to derive exact results from  Eq.(\ref{psi1}). In this case  $\varphi'$ and $\theta$ are constant, and since $\varphi'_-$ is negative whatever the sign of $\cos\theta$, the precession time reads $\tau_{s}=2\pi/\left\vert\varphi'_-\right\vert=-2\pi/\varphi'_-$.
 Combined with $\widetilde\Omega_3=t_0 \Omega_3=GI_1/I_3$,   
Eq.(\ref{psi1}) gives
\begin{eqnarray}
\frac{\Delta \Psi}{2\pi} & = & -\frac{I_1}{I_3}\frac{G}{\varphi'_-}{+}\cos\theta 
\label{psi3}
\end{eqnarray}
Using from Eq.(\ref{GPurePrecession})
\begin{eqnarray}
\nonumber
\frac{G}{\varphi'_{-}}=G^2\left(\frac{\varphi'_- \cos\theta}{G}-1\right)
\end{eqnarray}
 and using the slow precession velocity from Eq.(\ref{PurePrecessions})
\begin{eqnarray}
\nonumber
\frac{{\varphi'_-}\cos\theta}{G}=\frac{1}{2}\left(1-\sqrt{1+\frac{4\cos\theta}{G^2}}\right)
\end{eqnarray}
Eq.(\ref{psi3}) gives
\begin{eqnarray}
\frac{\Delta \Psi}{2\pi} & = & \frac{I_1}{I_3}G^2\left(\frac{1}{2}+\frac{1}{2}\sqrt{1+\frac{4\cos\theta}{G^2}}\right)+\cos\theta \\
&=& \frac{I_1}{I_3}G^2 + \cos\theta\left(1+\frac{I_1}{I_3}\right) - \frac{I_1}{I_3}\frac{\cos^2\theta}{G^2}\left(1-\frac{2\cos\theta}{G^2}+\frac{5\cos^2\theta}{G^4}+...\right)
\nonumber
\label{psi4}
\end{eqnarray}

This expression generalizes Eq.(\ref{DeltapsiG}) of Section III to the inertial regime for the pure precession. This is of course the same expression as that obtained for the spinning top in Ref.\cite{SpinTop}. In this framework, the first term $G^2 (I_1/I_3)$ of the expansion was the dynamical phase. The question that was discussed in Ref. \cite{SpinTop}, was about the nature of the second term $cos(\theta)$. There was an ambiguity about associating it to the dynamical phase or to the geometric phase. It appears below that, in the framework of the ``Bohr magneton'' approach used in section III-D for the magnetic monopole, the two first terms in the right hand side of Eq.(\ref{psi4}) are identical. Indeed, according to the II-D, we have $cos \theta = G^2 (I_1/I_3) = \Omega_3/\gamma H_z = t_s/t_f$ and Eq.(\ref{psi4}) reads : 

\begin{eqnarray}
\frac{\Delta \Psi}{2\pi}  &= & \frac{cos \theta}{2} \left(1 + \sqrt{1+4 \frac{cos \theta}{G^2} }\right)+\cos\theta \\
&= & cos \theta \left ( 2 +  \left(\frac{cos \theta}{G^2} \right) -    \left( \frac{cos \theta}{G^2} \right)^2 + 2    \left( \frac{cos \theta}{G^2} \right)^3 - 5    \left( \frac{cos \theta}{G^2} \right)^4 + ... \right)
\nonumber
\label{psi5}
\end{eqnarray}
The geometric phase $\Delta \Psi$ is a function of the precession angle $\theta$ and the slowness parameter $G$. Note that if we remove the dynamical angle $2cos \theta$, the developpement is a function of a single parameter $cos \theta / G^2$ only.

\subsection{The classical magnetic monopole for pure precession}
The generalization of the magnetic monopole Eq.(\ref{MonopoleMag}) is $B_{eff}= \frac{\Delta\Psi}{\pi R^2}$ so that

\begin{eqnarray}
B_{eff} &=& \frac{1}{R^2} \left [cos \theta \left(1 + \sqrt{1+4 \frac{cos \theta}{G^2} }\right)+2 \cos\theta \right ] \\
&= & \frac{2\cos\theta}{R^2}  \left ( 2 +  \left( \frac{cos \theta}{G^2} \right) -    \left(\frac{cos \theta}{G^2} \right)^2 + 2    \left( \frac{cos \theta}{G^2} \right)^3 - 5    \left( \frac{cos \theta}{G^2} \right)^4 + ... \right)
\nonumber
\label{MonopoleGene2}
\end{eqnarray}
This equation gives the influence of the inertia (i.e. the fast magnetic degrees of freedom) on the magnetic monopole, in the case of the pure precession.

\section{Conclusion}

Magnetization dynamics have been investigated beyond the usual assumption of the total separation of time scales between slow and fast magnetic degrees of freedom, for the adiabatic limit. We have exploited the analogy with the spinning top by pushing the mechanical model of the magnetic dipole beyond Gilbert's assumption. Fast degrees of freedom are introduced with the angular momentum $\vec L$ and its time variation (with non-zero first and second principal moment of inertia $I_1=I_2 \ne 0$). 

The problem is investigated from the viewpoint of the geometric phase which allows the magnetic monopole to be defined naturally. The effect of inertia is then taken into account, and an analytical expression is obtained in the case of the {\it pure precession}, for which the nutation vanishes.

In the case of pure precession with precession angle $\theta$, the calculation of the geometric phase shows that, beyond a dynamical phase of the form $2 cos \theta$, the Hannay angle is a simple function of the parameter $cos \theta /G^2$, where $G = t_s/(2 \pi t_0)$ is the slowness parameter (i.e. the ratio of the slow characteristic time of the precession over the fast characteristic time). 

The magnetic monopole (defined as the radial magnetic field produced from a punctual center), is derived directly from the geometric phase. In the usual case without inertia ($cos \theta /G^2 \rightarrow 0$), the Bohr magneton approach gives a very simple expression of the magnetic monopole as a function of the precession angle  $B_{eff} = 4 cos \theta /R^2$. In the case of pure precession, the correction due to the action of the fast degrees of freedom is given as a simple expression $B_{eff} = \frac{1}{R^2} \left [cos \theta \left(1 + \sqrt{1+4 \frac{cos \theta}{G^2} }\right)+2 \cos\theta \right ]$. Note that in an experimental context, the magnetic monopole $B_{eff}$ is constant because it is related to a given material, and the precession angle $\theta$ depends the parameter $G$.

This result suggests that the pure precession - i.e. the slow manifold for the dynamics of the magnetization \cite{SpinTop} -  should not be a purely formal concept, but could correspond to the actual motion of the magnetization for the ultrafast precession of the magnetization, that would correspond to the minimum power dissipated by the system (in comparison with the motion that includes nutation oscillations superimposed to the precession). This point should however still be clarified in further studies. 

\section{acknowledgments}
J.-E. W is grateful to Michael V. Berry for helpful comments.\\

\end{document}